\documentstyle[prl,aps,epsf,graphics,color]{revtex}

\def\m#1{$#1$}

\def\tr{\;{\rm tr}\;}
\def\sgn{\;{\rm sgn}\;}
\newcommand{\beq}{\begin{equation}}
\newcommand{\eeq}{\end{equation}}

\newcommand{\beqs}{\begin{eqnarray}}
\newcommand{\eeqs}{\end{eqnarray}}

\newcommand{\DOE}{This work was supported in part by U.S.Department 
of Energy grant No. DE--FG02-91ER40685}

\newcommand{\half}{\frac{1}{2}}
\newcommand{\eps}{\epsilon}

\begin{document}
\title{\bf\large The Anti-quark Distribution Function of the Baryon }

\author{
V. John\footnote{vjohn@pas.rochester.edu}, G. S. Krishnaswami\footnote{govind@pas.rochester.edu} and S. G. Rajeev\footnote{rajeev@pas.rochester.edu}\\
   {\it Department of Physics and Astronomy\\
 University of Rochester, Rochester, New York 14627} \\}
  
\maketitle

\abstract{ We derive the Deep Inelastic anti-quark distribution
in a baryon at a low value of \m{Q^2} using the variational 
principle of Quantum HadronDynamics, an alternative formulation
of Quantum ChromoDynamics. It is determined by a variational 
approach generalizing the ``valence'' quark approximation of 
earlier papers. We find that the ``primordial'' anti-quarks 
carry less than a percent of the baryon momentum.
In the limit of chiral symmetry and \m{N_c \to \infty},
we show that the anti-quark content of the proton vanishes at low \m{Q^2}.
}

\vspace{.3 cm}

{\it Keywords}: Structure Functions; Parton Model; Deep Inelastic
Scattering; Anti-quarks; Sea quarks; QCD; Skyrme model; Quantum
HadronDynamics.

{\it PACS }: 12.39Ki,13.60.-r, 12.39Dc,12.38Aw.

\vspace{.3 cm}

In previous papers \cite{2dqhd} one of us has outlined a way of
calculating the Deep Inelastic Structure functions of the baryon from
Quantum ChromoDynamics (QCD). After some approximations, ((i)
Dimensional Reduction to two dimensions, (ii) Ignoring transverse
gluon degrees of freedom) the theory reduces to two dimensional QCD,
which was transformed into a new form called Quantum HadronDynamics
(QHD). In this form the basic dynamical variable is  a color invariant
quatity \m{\hat
M(x,y)={1\over N_c}:[\chi_{\alpha}(x),\chi^{\dag \alpha}(y)]:},
(where \m{\chi,\chi^\dag} are the annihilation-creation operators of
quarks)  which can be thought of as the field operator of a meson.
The main advantage of this new point of view is that the (iii)
semi-classical approximation of QHD corresponds to the large \m{N_c}
limit of QCD, and so is capable of describing non-perturbative
phenomena such as the struture of hadrons. The baryon is a topological
soliton in this theory and its structure functions (within these
approximations) can be determined by a variational principle.
 In previous papers we made yet another approximation, 
(iv) the assumption of a factorized ansatz for the classical meson variable 
\m{M(x,y)=-2\psi(x)\psi^{*}(y)}, which 
corresponds to the valence quark approximation in the parton model. 

We have already discussed the consequences of relaxing some of these
simplifying assumptions. For example the effect of transverse momenta
(departure from two dimensionality: relaxing (i)) can be studied 
within perturbative QCD; indeed as emphasized by Altarelli 
and Parisi, this is the physical  meaning
of the usual DGLAP evolution equations \cite{dglap} for the structure
functions. The effect of reinstating transverse gluons (relaxing (ii))
 is to produce a slightly more involved two dimensional 
field theory, which we will
study in a separate paper. (This is important to derive the gluon
distribution functions of the baryon.) The effect of finite \m{N_c} is
(in the leading order) to restrict the range of values of the parton
momentum \cite{qcdp,ipm,istlect}. In this paper we will study the
departure from the factorized ansatz for \m{M(x,y)}; in other words,
 we will study the departure from the valence parton model. This is
the same as studying  the anti-quark content of the baryon. 

We will see that the probability of finding an anti-quark inside a
proton is quite small (\m{<1\%}) justifying the valence parton
approximations made in previous papers. Using a variational ansatz we
will obtain the anti-quark distribution functions. These can be used
as initial data for evolution in \m{Q^2} using the DGLAP equations, which take
into account  the perturbative corrections. That the initial
anti-quark content is quite  small, is consistent with the
phenomenological model of Gl\"uck and Reya \cite{grv}: we now have a
theoretical  derivation of this picture. However, we expect the initial
gluon distribution to be non-zero. There are other approaches to studying parton distribution functions, see for example Ref. \cite{brodskyetal}.

Let us begin by summarizing the large \m{N_c} limit (which is the
classical limit) of QHD. The dynamical variable is a complex valued
function \m{M(x,y)} of two space-time points \m{(x,y)} lying along 
 a null-line. This variable satisfies 
\m{
	M^*(x,y)=M(y,x)
}
so that we can regard it as the integral kernel of a hermitean 
 operator on \m{L^2(R)}. (For technical reasons we assume that this
operator is Hilbert-Schmidt; i.e., \m{\int |M(x,y)|^2dxdy<\infty}.) 
Moreover, it satisfies the non-linear constraint \m{[\eps+M]^2=1} 
where the operator \m{\eps} is the celebrated Hilbert transform 
operator with \m{\eps^2=1} and the integral kernel \m{\eps(x,y)=\int
\sgn(p)e^{ip(x-y)}{dp\over 2\pi}={i\over \pi}{\cal P}{1\over x-y}}. 
This constraint can be understood as a consequence of the Pauli principle 
for fermions as explained in \cite{istlect}. The static solutions 
of the theory are then the minima of the energy functional 
\beq
{E(M)\over N_c} ={-{1\over 2}} \int \tilde M(p,p)\half\left[p+{\mu^2\over p}\right]
{dp\over 2\pi}
+{\tilde
g^2\over 8}\int dxdy |M(x,y)|^2\half|x-y|
\eeq
subject to the above constraints.
The first term is just the kinetic energy in null co-ordinates \cite{2dqhd};
the second is the potential energy induced by the
longitudinal gluon fields.
 (Recall that the linear potential \m{\half|x-y|} is the
Fourier transform of the gluon propagator \m{1\over q^2} in two 
space time dimensions). The parameter \m{\tilde g \sim \Lambda_{QCD}} 
determines the strength of the strong interactions; also,
\m{\mu^2=m^2-{{\tilde g}^2\over \pi}} is related to the current quark
mass \m{m} through a finite renormalization. We will be mainly 
interested in the case \m{m<<\tilde g} which corresponds to the limit
of chiral symmetry.  It has been  shown elsewhere \cite{istlect} that
subject to the above constraints, the energy \m{E(M)} is positive: the
constraints being crucial for this conclusion. A Lorentz invariant
form of the above variational principle is to minimize the invariant
mass-squared \m{{\cal M}^2} rather than energy. Since the null  momentum of a
configuration is \m{{P\over N_c}=-\half \int p\tilde M(p,p){dp\over 2\pi}} we get
\beq
{{\cal M}^2 \over N_c^2}=\left[-\half \int p\tilde M(p,p){dp\over
2\pi}\right]\left[{-{1\over 2}} \int \tilde M(p,p){\mu^2\over
2p}{dp\over 2\pi}+{\tilde
g^2\over 8}\int dxdy |M(x,y)|^2\half|x-y|\right]
\eeq 

The quantity \m{B=-\half\tr M=-\half\int M(x,x)dx} is an integer, 
a topological invariant of the configuration as shown in \cite{2dqhd}. 
If we reexpress it in terms of the quark fields we can see that this 
is just the baryon number. Thus, a baryon is a topological soliton 
in this picture. We can get the structure functions of the baryon 
by minimizing the energy functional \m{E(M)} subject to the above constraints.
We have developed a method \cite{istlect} to solve this problem: a
variant of the steepest descent method that takes into account the
non-linear constraint. However, this method is computationally 
intensive. A method based on a variational ansatz that builds on our previous
results on the separable ansatz gives as good results while being much
simpler computationally. We report on the results of this variational
approach in this paper.

The separable ansatz,
\m{\tilde M(p,q)=-2\tilde\psi(p)\tilde\psi^*(q)} satisfies the
constraint if the vector
\m{\psi\in L^2(R)} is of unit length and is non-zero only for positive
momentum.  This is easily verified by substitution.

In the limit of chiral symmetry \m{m=0}, and
\m{N_c\to \infty} the {\it exact } minimum of the functional \m{{\cal
M}^2}   is of the separable form, with \m{\tilde\psi(p)\sim
e^{-{p\over \tilde g}}}. To see this we first  derive an integral
equation for the extremum
of \m{{\cal M}^2}} by varying  with respect to \m{M}, respecting the
constraint. In operator language this
equation is \m{\left[\eps+M,{\delta{\cal M}^2\over \delta
M}\right]=0}, and can be converted to a nonlinear integral equation.
 By direct computation, the separable ansatz with \m{\tilde\psi(p)\sim
e^{-{p\over \tilde g}}} can then be verified to be an exact solution of
this equation, in the limit \m{m=0}. (Details will be given in a
longer paper.) Moreover, \m{{\cal M}^2} is zero for this solution. 
Thus the  departure from the valence parton  picture is determined 
by the dimensionless ratio \m{{m^2\over \tilde g^2}} which quantifies
chiral symmetry breaking. (As we showed in Ref. \cite{ipm} 
the leading effect of finite \m{N_c} is not to depart from the valence parton
picture but rather to constrain the range of momenta of the partons.)
 
Thus we should expect the `primordial' anti-quark distribution in the
proton to be small: the up and down quarks have current quark masses small in
comparison to \m{\Lambda_{\rm QCD}}, which means that 
\m{{m^2\over \tilde g^2}<<1} as well.

The mathematical advantage of the separable ansatz is that it `solves'
the  nonlinear  constraint on \m{M}: more precisely, it replaces it
with the condition that \m{\psi} is of norm one. In the same spirit,
consider the confguration 
\m{M=\sum_{a,b=1}^r\xi_b^a\psi_a\otimes \psi^{\dag b}}. Here we choose
\m{\psi_a} to be a set of \m{r} orthonormal eigenvectors of the operator
\m{\eps}; i.e., \m{\eps\psi_a=\eps_a\psi_a}, \m{\eps_a=\pm 1}. This
implies that the operator \m{M} is of rank \m{r}: the special case
of rank one is just the separable ansatz above. This ansatz
will satisfy the constraint on \m{M} if the \m{r\times r} matrix
\m{\xi} is hermitean and satisfies the  constraint
\m{\xi_a^b\xi_b^c+[\eps_a+\eps_c]\xi_a^c=0}: a `mini' version of the
constraint on \m{M}. Moreover, the baryon number is \m{B=-\half {\rm tr}\;
M= -\half tr \xi}. In the special case of rank one, we have simply
\m{\xi=-2}. By choosing a large enough value of \m{r} this ansatz can
produce as general a configuration in the phase space as needed: such
configurations form a dense subset of the phase space. 

The simplest configuration of baryon number one that departs from the
separable ansatz is of rank three. We will find that in physically
interesting situations, even this departure is very small, so we do not
need to consider configurations of higher rank.

By a choice of basis among the \m{\psi_a},  we can always
bring a rank three configuration of baryon number one  to the form 
\m{M=-2\psi\otimes \psi^\dag + 
2\zeta_-\big\{\zeta_-[\psi_-\otimes\psi_-^\dag -\psi_+\otimes\psi_+^\dag] +
\surd[1-\zeta_-^2][\psi_-\otimes\psi_+^\dag +\psi_+\otimes\psi_-^\dag]\big\}}
where \m{\psi_-,\psi,\psi_+} are three vectors in \m{L^2(R)} satisfying
\m{\eps\psi_-=-\psi_-,\quad  \eps\psi=\psi,\quad 
\eps\psi_+=\psi_+, ||\psi_-||^2=||\psi||^2=||\psi_+||^2=1,\quad <\psi,\psi_+>=0.}
The conditions \m{<\psi_-,\psi>=<\psi_-,\psi_+>=0} are then automatic.
The parameter \m{0\leq \zeta_-\leq 1} measures the deviation from 
the rank one ansatz and hence, the anti-quark content of the baryon. 
For example, baryon number is given by 
\m{B=\int_0^\infty\left\{|\tilde\psi(p)|^2+
\zeta_-^2\left[|\tilde\psi_+(p)|^2-|\tilde\psi_-(-p)|^2\right]\right\}
{dp\over 2\pi}}. The wavefunctions \m{\psi,\psi_+} both describe
quarks and their orthogonality can be interpreted as a consequence of
the Pauli principle. \m{\psi} describes ``valence'' quarks while \m{\psi_+}
is the wave function of the ``sea'' quarks. Since \m{\psi_-} 
contributes with a negative sign to the baryon number, it 
describes anti-quarks. From our previous result we expect \m{\zeta_-} to
vanish as \m{{m^2\over \tilde g^2}\to 0}.

We can substitute this ansatz into the energy \m{E(M)} or 
\m{{\cal M}^2} and derive integral equations for the 
minimization. However, in keeping with the
spirit of the variational ansatz, we can simplify the problem by
assuming first  some simple functional
forms for the functions \m{\tilde\psi,\tilde\psi_\pm}. The form
of the exact solution suggests the choice
\m{\psi(p)=C\left({p\over \tilde g}\right)^a e^{-b{p\over \tilde g}},
\psi_+(p)=C_+\left({p\over \tilde g}\right)^a\left[{p\over \tilde
g}-C_1\right] e^{-b{p\over \tilde g}}} for \m{p>0} and
\m{\tilde\psi_-(p)=\tilde\psi(-p) } for \m{p<0}. (For other ranges of
\m{p} these functions must vanish.) The parameter \m{C_1} is 
determined by the orthogonality condition while \m{C,C_+} are fixed by
the normalization conditions. The variational parameter \m{b} 
determines the reference frame. The Lorentz invariant quantity
\m{{\cal M}^2} is independent of \m{b}. Thus the variational
principle will determine \m{a} and \m{\zeta_-} and hence the wavefunctions.

The rest of the calculation is a straightforward evaluation of 
the energy integrals and then their
minimization. (We use the symbolic package {\it Mathematica} for some 
of the computations, most of which can be done analytically. Some details 
are provided in \cite{istlect}). We have done the calculation and 
shown that for physically interesting values of \m{{m^2\over \tilde g^2}}
 (\m{\sim ({m_{u,d}\over\Lambda_{QCD}})^2\sim .001}), the
parameter \m{\zeta_-} is quite small. We present the results in the
figures which show the small effects of deviations from the separable ansatz 
(i.e. the effects of anti-quarks) and from chiral symmetry. Finally, 
the effect of finite \m{N_c} is (in the leading order)
to restrict  the maximum value of parton momenta. We have already 
studied this correction in the case of the separable ansatz and find 
it to be small \cite{qcdp,ipm,xf3}. In the case of anti-quarks, we 
establish that they carry less than a percent of baryon momentum 
in the \m{N_c \to \infty} limit and therefore, corrections due to 
finite \m{N_c} are less relevant. They will be addressed in a longer paper.

Thus, we have derived the ``primordial'' anti-quark 
distribution function of the proton 
by a series of approximations from QCD. We have an explanation of 
why it is small in comparison to the valence quark distribution at 
the low initial value of \m{Q_0^2 \sim 0.4 GeV^2} \cite{xf3}. The 
anti-quark distribution is in fact zero in the limit of chiral 
symmetry and when \m{N_c \to \infty}, while deviations are small.
This justifies the valence parton
approximations made in earlier papers \cite{ipm,qcdp,xf3}. 
It is possible to compare our prediction with experimental data: 
there is a specific combination of deep inelastic structure functions 
that describes anti-quarks \cite{cteqmrst}. To make a comparison, we need to 
evolve our distribution from \m{Q_0^2}, according 
to the DGLAP equations. However, it is necessary to know the
initial gluon distribution in order to solve the evolution equations. 
We will study the gluon distribution in a later paper and subsequently 
return to this issue.

Acknowledgement: \DOE.

\begin{center}
\scalebox{0.5}{\includegraphics{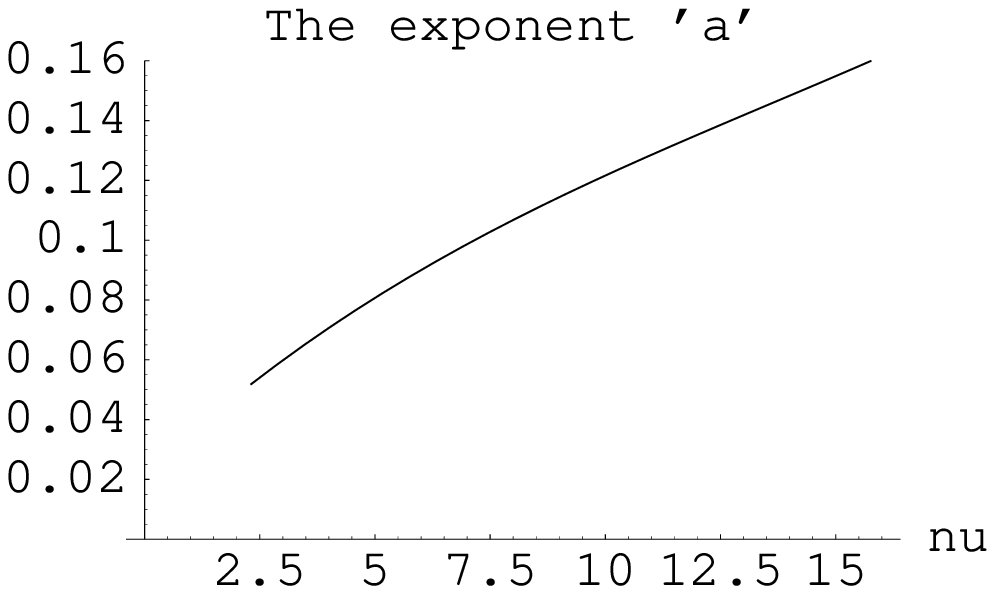}}
\scalebox{0.5}{\includegraphics{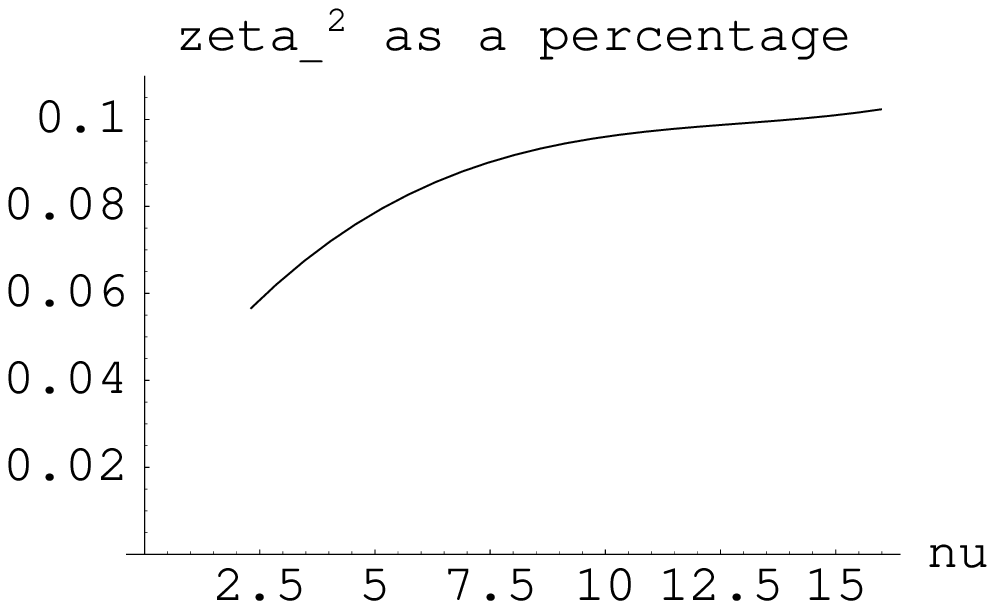}}
\end{center}

Figure 1: Variational estimates for (a) the exponent '\m{a}' and (b) 
\m{\zeta_-^2} as a percent. They are plotted as functions of 
\m{nu = 1000*{m^2\over\tilde{g}^2}}. The exponent \m{a} and the anti-quark 
content \m{\zeta_-^2} go to zero for small current quark masses.

\begin{center}
\scalebox{0.5}{\includegraphics{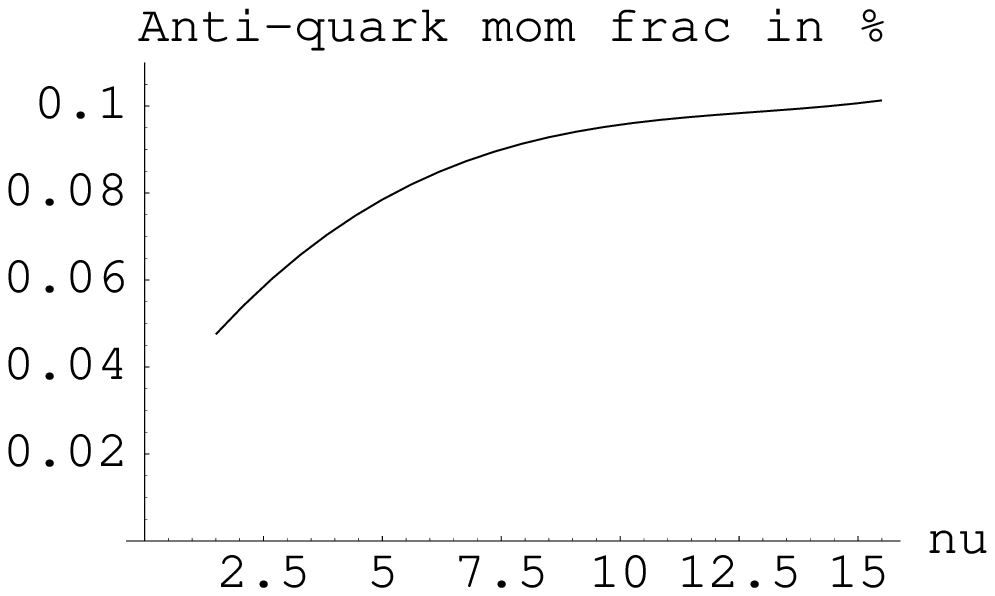}}
\scalebox{0.5}{\includegraphics{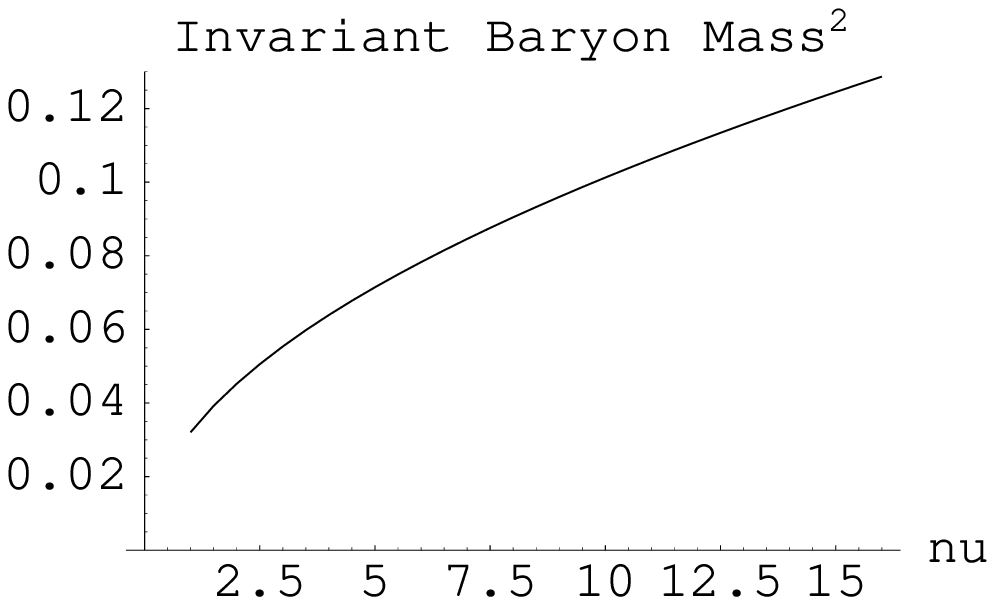}}
\end{center}

Figure 2: (a) Variational estimate for the fraction of fermion
momentum of the baryon carried by anti-quarks. It is plotted as a percent
as a function of \m{nu}. The ``primordial'' 
anti-quarks carry less than a percent of the portion of baryon momentum 
shared between quarks and anti-quarks.
(b) Variational upper-bound on the invariant \m{Mass^2},
(\m{{{\cal M}^2\over\tilde{g}^2N_c^2}}), of the baryon in the two-dimensional 
approximation, plotted as a function of \m{nu}. 
In the limit of chiral symmetry, we recover the exact exponential
solution with \m{{{\cal M}^2\over\tilde{g}^2N_c^2} = 0}, \m{\zeta_- = 0} 
and \m{a=0}.

\begin{center}
\scalebox{0.5}{\includegraphics{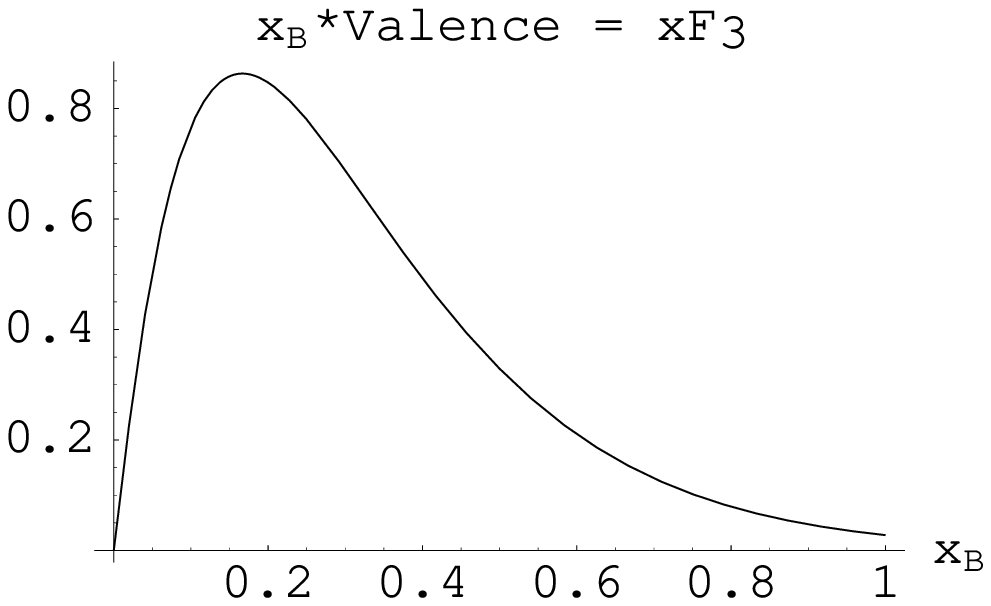}}
\scalebox{0.5}{\includegraphics{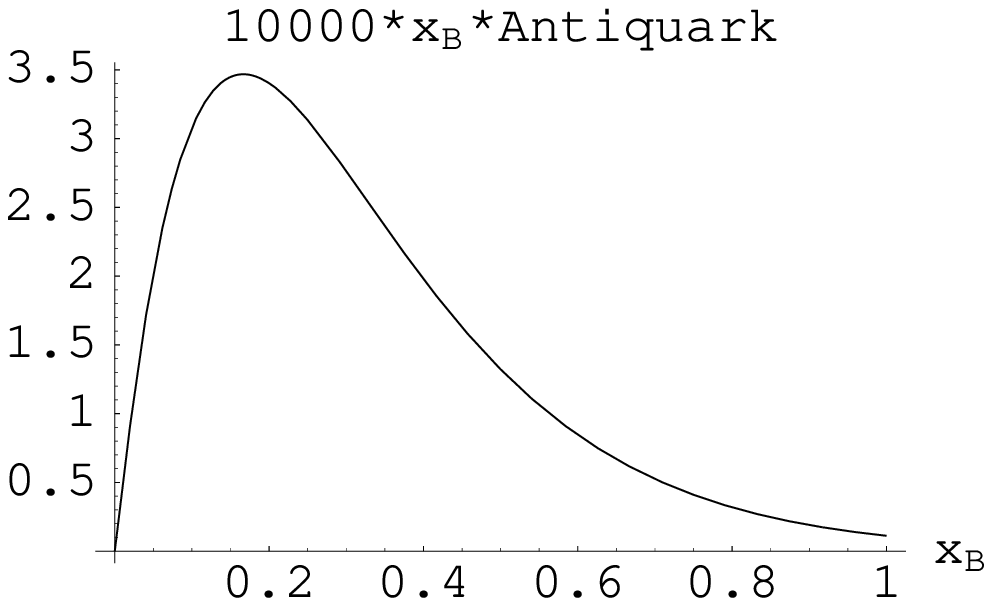}}
\scalebox{0.5}{\includegraphics{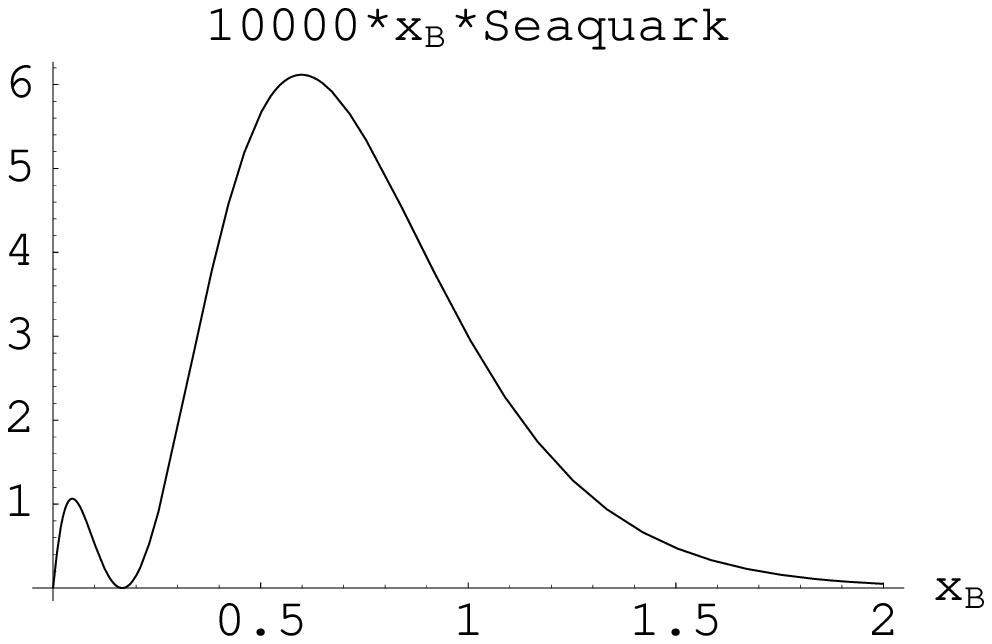}}
\end{center}

Figure 3: (a) The ``valence'' quark (\m{x_B*|\psi|^2}),
(b) anti-quark (\m{x_B*|\psi_-|^2}) and (c) ``sea'' quark (\m{x_B*|\psi_+|^2})
distributions plotted as a function of momentum fraction 
\m{x_B ={p\over{P}}} at low \m{Q^2} (\m{\sim 0.4 GeV^2}, see Ref.\cite{xf3}). 
The exponential tails beyond \m{x_B = 1} are an artifact of the large-\m{N_c} limit. They are plotted for a small value 
of current quark mass 
(\m{{m^2\over\tilde{g}^2}~\sim~({m_{u,d}\over\Lambda_{QCD}})^2\sim~.001}) in 
the reference frame in which the mean baryon momentum, \m{P} is 1. The
fermions are assumed to carry \m{f= \half} the mean baryon momentum. The 
rest is carried by gluons \cite{xf3,cteqmrst}. The node in the ``sea'' quark 
wavefunction is because it is required to be 
orthogonal to the ``valence'' quark wavefunction by the Pauli principle.
The valence quark distribution shown above (\m{x_B*} Valence \m{= xF_3}), 
though calculated in the limit
 \m{N_c \to \infty} agrees well with the distribution obtained after
 taking into account the leading \m{{1\over{N_c}}} corrections 
and also with experimental measurements of the neutrino structure function 
\m{x_B F_3(x_B,Q^2)} when evolved to higher values of \m{Q^2} \cite{xf3}.

\end{document}